\documentclass[letterpaper, 10 pt, conference]{ieeeconf}  
\usepackage{graphicx}
\usepackage{amsmath}
\usepackage{amssymb}
\usepackage{multicol}

\IEEEoverridecommandlockouts                              

\title{\LARGE \bf
 Optimal adaptive control of a knee joint exoskeleton for lower limb functional rehabilitation
}

\author{Maria-Sara-Nour Sadoun$^{1}$
, Fouad Yacef$^{2}$
\thanks{$^{1}$ Maria-Sara-Nour Sadoun is with Ecole Nationale Polytechnique , Algiers, Algeria
 {\tt\small maria\_sara\_nour.sadoun@g.enp.edu.dz}}%
 \thanks{$^{2}$ Fouad Yacef is with Centre de D\'eveloppement des Technologies Avanc\'ees (CDTA), Algiers, Algeria,
{\tt\small fyacef@cdta.dz}}%
}

\begin{document}
This work has been submitted to the IEEE for possible publication. Copyright may be transferred without notice, after which this version may no longer be accessible.

\maketitle
\thispagestyle{empty}
\pagestyle{empty}

\begin{abstract}
Lower limb exoskeleton robots hold great potential for rehabilitation, movement assistance, and strength augmentation. Design control to guarantee optimal needed assistance is still a challenge considering the pathological variances between patients. In this paper, we proposed an optimal adaptive control scheme based on Particle Swarm Optimization (PSO) Algorithm. The proposed controller is based on a well-known dynamic model of the knee joint exoskeleton, and the optimization algorithm is used to minimize a square error fitness function, which quantifies tracking performances. Control parameters are tuned respecting some nonlinear constraints for step response of the system and boundaries constraints. Numerical simulation results are presented to show the validity and the high performances of the proposed approach.

\vspace{0.1cm}
\textbf{\textit{Keywords---} Particle Swarm Optimization (PSO), Adaptive control, Constrained optimization, Lower limb exoskeleton} 
\end{abstract}

\section{INTRODUCTION}
Exoskeleton robots enable movement assistance and neural rehabilitation through instant post-stroke therapy as well as walking mobility. Repetitive, highly controlled tasks and accelerated training can be ensured by robotic rehabilitation, which, compared with medical therapy, reduce the bother of medical staff and deliver a significant evaluation of human activities through motion and forces measurements \cite{huo2016lower} \cite{young2017state}.

Recently, lower-limb
exoskeletons attract more and more interest in the research community, both in engineering
and medical fields. This growing interest is due to the wide range of applications that can be
addressed in functional rehabilitation  \cite{rifai2017toward} \cite{ajjanaromvat2018trajectory}, strength improvement \cite{cenit2020design}, and walking assistance \cite{peng2020datab} \cite{peng2020optimal} \cite{zhang2019hierarchical}. In the fields of rehabilitation and walking assistance, exoskeletons are
designed for assisting paraplegic patients
whose lower-body is disabled \cite{huo2016lower}, hemiplegic patients
with one side of the body weakened \cite{alshatti2019design} \cite{peng2020data} and monoplegic patients suffering from one-limb paralysis restrictions \cite{caulcrick2020model} \cite{sun2020reduced}.
While it has been remarked that stroke became a global healthcare
problem, many research studies on exoskeleton focused on
walking assistance or rehabilitation case for monoplegic.

In more recent studies on rehabilitation using lower-limb exoskeletons to help stroke patients relearn motion patterns, researchers proposed three types of intensive exercises: passive, active-assist, and active
\cite{peng2020datab}, adapted according to the patients’ state. While exoskeleton robot assistance is necessary for passive exercises, active-assist exercises involve the patient in the movement, and robots deliver appropriate assistance by sensing the patient’s physical limitations.

Many control approaches have been developed to drive the
lower limb exoskeleton for a different mode of assistance. One
can cite, data-driven control \cite{peng2020datab} \cite{peng2020data} \cite{tu2020data}, adaptive control \cite{rifai2012adaptive} \cite{rifai2016augmented}, model-free based control \cite{talbi2013rbf} \cite{zhang2018model}, bounded control \cite{rifai2013nested} \cite{rifai2011bounded}, model predictive control  \cite{caulcrick2020model} and sliding mode control technique \cite{mohammed2016nonlinear} \cite{huo2019impedance}. A comparative study between different control techniques for lower limb orthosis in different operating conditions was provided in \cite{roula2019control}. The authors in \cite{caulcrick2020model} proposed a control architecture for lower limb assistance robots based on a fuzzy logic system (FLs) for modes of assistance identification and a model predictive controller (MPC). Muscle activity is sensed by electromyography (EMG), enabling MPC to predict human torque and FLS to select assistance mode. In \cite{peng2020datab} \cite{peng2020data}, to handle assistance control problems in lower limb exoskeletons aimed at helping hemiplegic patients walking, the authors developed a reinforcement learning control strategy. The lower limb exoskeleton systems, where patient legs are included, are modeled within a leader-follower framework. The walking assistance control problem is transformed into an optimal control problem and then solved by a policy iteration (PI) algorithm. A modified version of Particle Swarm Optimization (PSO) algorithm called Adaptive PSO (APSO) was proposed in \cite{soleimani2020adaptive} to tune a PID controller gain for lower-limb human exoskeleton control.\\

This paper presents the tuning of adaptive controller parameters using PSO optimization algorithm under the considered system’s physical boundaries constraints and nonlinear constraints for step response. The dynamic model of the lower limb exoskeleton is described in SECTION I followed in SECTION II with the development of the adaptive controller that ensures the asymptotic stability of the closed-loop system. An objective function is
formulated in terms of square error between the actual and the desired angular positions and velocities of the shank. The main objective is to design an optimization procedure using a meta-heuristic optimization algorithm to optimize controller parameters under nonlinear constraints.\\
\section{SYSTEM'S MODELING}

The system considered is composed of the human shank and the embodied actuated orthosis and has one degree of freedom. The orthosis subject of this research project is a mechanical device designed to fit perfectly the subject’s leg it is fixed to using straps and thus matches its geometry to ensure their coupling and synchronous motion when driven by the actuator  \cite{rifai2011bounded}. The designed device is subject to the human torque delivered by the thigh muscles and to the actuator torque computed through the chosen adaptation control aimed to track the desired trajectory through flexion-extension movements for rehabilitation purposes.\\

The system considered is composed of the human shank and the embodied actuated orthosis and has one degree of freedom. To model the shank-foot-exoskeleton system, the rotational dynamics can be expressed as follows, and when defining its dynamics, we define two inertial frames: one fixed $\mathcal{A}\left(\vec{x}_{A}, \vec{y}_{A}, \vec{z}_{A}\right)$ and one attached to the shank foot at the knee articulation $\mathcal{B}\left(\vec{x}_{B}, \vec{y}_{B}, \vec{z}_{B}\right)$. Both inertial frames $\vec{y}$ axis coincide, and the system is in rotation about the knee joint with an angle $\theta$ about the axis $\vec{y}$ and the angle is defined relatively to the thigh as the direction of the shank \cite{rifai2016augmented}.\\
\begin{figure}
\centering
\includegraphics[scale= 0.35]{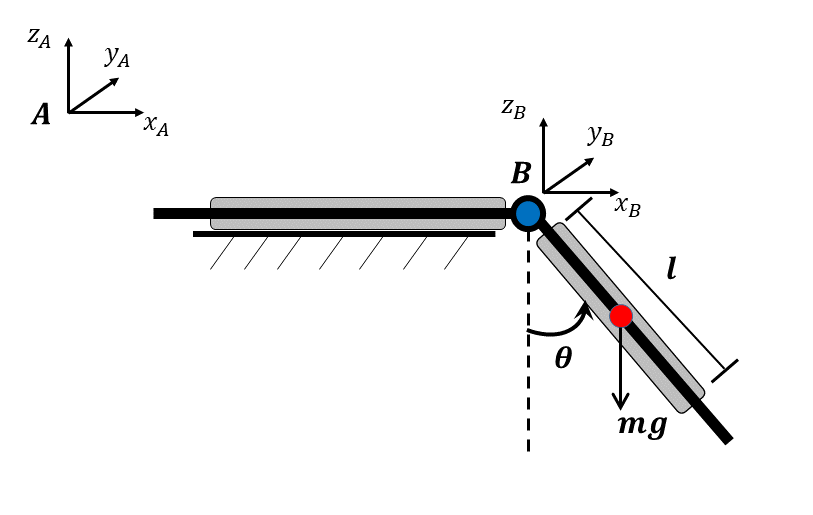}
\caption{Human leg embodying the orthosis: Fixed and Shank frames.}
\end{figure}

We define $\theta$ and $\Dot{\theta}$ respectively as the angular position and velocity of the shank relative to the thigh. Dividing the system to its two components, the leg and orthosis, will be studied, and their model developed similarly. This dissociation aims at facilitating the analyses of the applied forces.\\

The kinetic and gravitational energies of each component are defined as: $E_{k_i}=\frac{1}{2}I_i\Dot{\theta}^2$ and $E_{g_i}=m_igl_i(1-sin\theta)$ with $i \in \{1,2\}$ where $I_i$ is the inertia of the component, $m_i$ the mass, $l_i$ the distance from the knee joint to the component center of gravity and $g$ is the gravity acceleration.The Lagrangian is defined for each component of the studied system as: $\mathcal{L}_l=E_{k_i}-E_{g_i}$.

When deriving it, the dynamics of each component is defined as: $I_i\ddot{\theta}+m_igl_icos\theta=\Gamma_{ext_i}$, $\Gamma_{ext_i}$ being the external torque acting on the component, including the control torque delivered by the on-board actuator $\Gamma$ and the friction torques $\Gamma_f$.
The friction torque encompasses the solid and viscous frictions: $\Gamma_f=-C_{si}sign(\Dot{\theta})-C_{vi}\Dot{\theta}$ where $C_{si}$ and $C_{vi}$ are respectively the solid and viscous frictions coefficients and $sign(.)$ is the classical sign function. The advantage noted from the chosen friction model is the linearity in the previous coefficients.

The human torque noted $\Gamma_h$ is developed by the subject given the necessary effort to control his muscles. It is considered bounded given that the user during rehabilitation is considered to deliver a limited effort: $|\Gamma_h| \leq \alpha_h$.
The dynamics of the system (orthosis and human shank) is, therefore, given as:
\begin{equation}
    I\Ddot{\theta}=-\Gamma_g cos\theta+\Gamma+\Gamma_h-C_s sign(\Dot{\theta})-C_v\Dot{\theta}
    \label{eq1}
\end{equation}

Within the assistance strategies, $\Gamma_h$ is defined as contributing to the movement, and conversely, within the resistive rehabilitation, $\Gamma_h$ is defined as a resistive torque as it opposes the tracking of the desired trajectory.

\section{ADAPTIVE CONTROL DESIGN}

Adaptive control is widespread in rehabilitation and assisting devices, given that its concept allows the compensation of changes in the human-exoskeleton system dynamics. The law is based on a modulation of the control parameters function of the tracking performance and
consists on the continuous computing of the said parameters in order to reach the best convergence to the reference trajectory. The desired angle,
angular velocity and acceleration are respectively denoted $\theta_d$, $\Dot{\theta}_d$, $\Ddot{\theta}_d$ and the errors on the current state variables are expressed as follows:  $\Tilde{\theta}=\theta-\theta_d$, $\Dot{\Tilde{\theta}}=\Dot{\theta}-\Dot{\theta}_d$, $\Ddot{\Tilde{\theta}}=\Ddot{\theta}-\Ddot{\theta}_d$. Define also the variable -also known as sliding surface slope- $\Dot{s}=\Ddot{\Tilde{\theta}}+\gamma\Dot{\Tilde{\theta}}$ with $\gamma$ a positive scalar parameter.

\textbf{Assumption 1:} The current and desired states up to the third derivative ($\theta,$ $\Dot{\theta}$, $\Ddot{\theta}$, $\theta^{(3)}$, $\theta_d,$ $\Dot{\theta}_d$, $\Ddot{\theta}_d$, $\theta^{(3)}_d$) are bounded.

Denote $\hat{I}$, $\hat{\Gamma}_g$, $\hat{C}_s$, $\hat{C}_v$ the estimated system's parameters as defined in the modelling. The control torque is defined as \cite{rifai2012adaptive}:
\begin{equation}
    \Gamma=\hat{I}(\Ddot{\theta}_d-\gamma\Dot{\Tilde{\theta}})+\hat{C}_s sign(\Dot{\theta})+\hat{C}_v\Dot{\theta}-\kappa s+\hat{\Gamma}_g cos\theta
\end{equation}
with $\kappa$ a positive parameter and the dynamics of the adaptive laws are given as:
\begin{equation}
\Dot{\hat{I}}=-\eta_1(\Ddot{\theta_d}-\gamma\Dot{\Tilde{\theta}}) s
\end{equation}
\begin{equation}
\Dot{\hat{C}}_s=-\eta_2sign(\Dot{\theta}) s
\end{equation}
\begin{equation}
\Dot{\hat{C}}_v=-\eta_3\Dot{\theta} s
\end{equation}
\begin{equation}
\Dot{\hat{\Gamma}}_g=-\eta_4cos\theta s
\end{equation}

with $\eta_1$, $\eta_2$, $\eta_3$, $\eta_4$ are positive scaling parameters.

\textbf{Remark:} Note that as previously mentioned, the system's parameters are in this method in-line computed in order to track best the desired trajectory. Therefore, the non-convergence of the estimated parameters at stability to the real established values is an common eventuality and not contradictory with the right unfolding of the method.

When deriving the $s$ mathematical expression $\Ddot{\theta}=\Dot{s}-\gamma \Dot{\theta}$ and replacing (2) in (1), results in the dynamics of the closed loop system:
\begin{equation}
    I\Dot{s}=\Gamma_h-\kappa s-\Tilde{I}(\Ddot{\theta}_d-\gamma\Dot{\Tilde{\theta}})-\Tilde{C}_s sign(\dot{\theta})+\Tilde{C}_v\Dot{\theta}
\end{equation}
where $\Tilde{I}=I-\hat{I}$, $\Tilde{C}_s=C_s-\hat{C}_s$, $\Tilde{C}_v=C_v-\hat{C}_v$ and $\Tilde{\Gamma}_g=\Gamma_g-\hat{\Gamma}_g$.

Two modes of assistance are faced. The first one is the case of a passive wearer; hence no human effort is delivered where the asymptotic stability of the system equilibrium will be proved. The second case will concern an active subject capable of delivering a bounded human torque. Our study considers the first assistance mode of passive wearer.\\

\textbf{Proposition 1:} The system considered is described in (1) with  $\Gamma_h=0$. The derivative of the position variable corresponding to the desired trajectory is supposed bounded up to the third derivative. The asymptotic stability ensured in this case is at $x=(\Tilde{\theta}, \Dot{\Tilde{\theta}})=(0,0)$ with a domain of attraction $(-\pi, \pi)\times\mathrm{R}$. This result is synonym to a decreasing Lyapunov function. Since all concerned signals were previously considered bounded, this translates into the boundedness of the second derivative of the Lyapunov function $V$ . This assessment implies the uniform continuity of $\ddot{V}$. By means of the Barbalat Lemma, the asymptotic convergence of the state with the domain of attraction mentioned beforehand. Explicitly, the knee-joint exoskeleton coupled with the human leg is asymptotically stable, tracks the desired trajectory and is bounded.

    
    \textit{Proof:} The positive Lyapunov function is beforehand defined as:
    \begin{equation}
        V=\frac{1}{2}Is^2+\frac{1}{2\eta_1}\Tilde{I}^2+\frac{1}{2\eta_2}\Tilde{C}_s^2+\frac{1}{2\eta_3}\Tilde{C}_v^2+\frac{1}{2\eta_4}\Tilde{\Gamma_g}^2+\kappa\gamma{\tilde{\theta}}^2
    \label{eq8}
    \end{equation}
    Considering the system's parameters constant and replacing (3, 4, 5, 6):

$$
\begin{aligned}
\dot{V} &=s I \dot{s}+\frac{1}{\eta_{1}} \tilde{I} \dot{\bar{I}}+\frac{1}{\eta_{2}} \tilde{C}_s \dot{\dot{C}}_s+\frac{1}{\eta_{3}} \tilde{C}_v \dot{\bar{C}}_v+\frac{1}{\eta_{4}} \tilde{\Gamma}_{g} \dot{\tilde{\Gamma}}_{g} +2\kappa\gamma \Dot{\tilde{\theta}}\tilde{\theta}\\
&=-\kappa (\tilde{\theta}^{2}+\gamma^2\dot{\tilde{\theta}}^2) \leq 0
\end{aligned}
$$
This result is synonym to a decreasing Lyapunov function. Since all concerned signals were previously considered bounded, this translates into the boundedness of the second derivative of the Lyapunov function $\Ddot{V}$. This assessment implies the uniform continuity of $\dot{V}$. By means of the Barbalat Lemma, the asymptotic convergence of the state with the domain of attraction mentioned beforehand. Explicitly, the knee-joint exoskeleton coupled with the human leg is asymptotically stable, tracks the desired trajectory and is bounded.

\textbf{Remark:} As explained earlier, the adaptive laws do not
necessarily converge to the actual values of the system. Rather,
the core goal of the control law is the sole convergence and
stability of the system when tracking the desired trajectory
established prior to the rehabilitation process.

\section{OPTIMIZATION PROCEDURE}
\subsection{Particle swarm optimization}
\begin{figure}[ht]
\centering
\includegraphics[scale= 0.7]{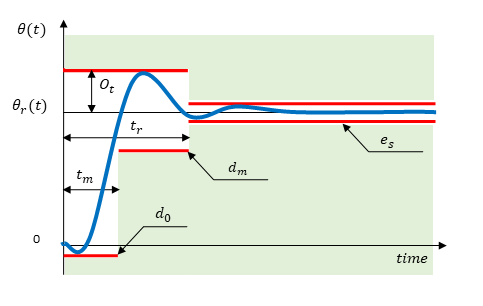}
\caption{Constraints on the step-response}
\end{figure}
Particle swarm optimization is a stochastic search method inspired by the coordinated motion of animals living in groups. The common goal of all group members is to find the most favorable location within a specified search space. The PSO algorithm searches in parallel using a group of individuals similar to other population-based heuristic optimization techniques. PSO technique searches using a population of particles corresponding to individuals. Each particle represents a candidate solution to the problem at hand. In a PSO system, particles change their positions by “flying” around in a multidimensional search space. Particle in a swarm adjusts its position in search space using its present velocity, own previous experience, and that of neighboring particles. Therefore, a particle uses the best position
encountered by itself and that of its neighbors to steer toward an optimal solution \cite{yacef2013pso}.
\begin{figure}[ht]
\centering
\includegraphics[scale= 0.5]{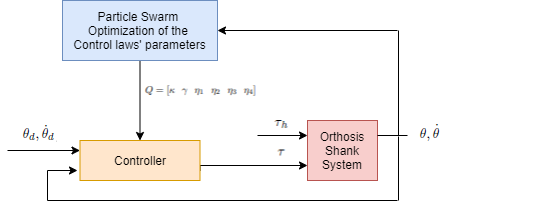}
\caption{Optimal adaptive Control Scheme}
\end{figure}

The PSO algorithm formulation adopted in this study is given by the following equations \cite{ebbesen2012generic}
\begin{equation}
\begin{cases}
    v_i^{k+1} = w^k v_i^k + \alpha_1\beta_{1,i}(P_i-x_i^k) + \alpha_2\beta_{2,i}(G_i-x_i^k)\\
    x_i^{k+1} = x_i^k + v_i^{k+1} 
\end{cases}
\end{equation}
The vectors $x_i^k$ and $v_i^k$ are the current position and velocity of the $i$-th particle in the $k$-th generation. The swarm consists of $N$ particles, i.e. $i={1,\ldots,N}$. Furthermore, $P_i$ is the personal best position of each individual and $G$ is the global best position observed among all particles up to the current generation. The parameters $\beta_{1,2} \in [0,1]$ are uniformly distributed random values and $\alpha_{1,2}$ are acceleration constants. The function $w$ is the particle inertia which gives rise to a certain momentum of the particles.
\begin{figure}[ht]
\centering
\includegraphics[scale= 0.5]{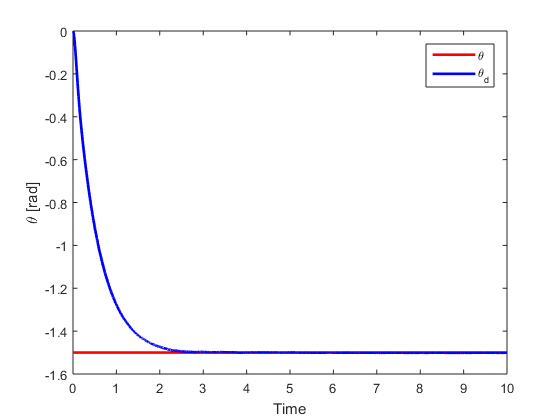}
\caption{Actual and desired  angular position of knee joint $\theta$}
\end{figure}
\subsection{Controller parameters tuning}
The tuning parameters vector can be defined based on the system knowledge and the control objective. In the proposed PSO tuning procedure for lower limb exoskeleton controller, the tuning vector Q is composed by the parameters of our controller $\kappa, \gamma, \eta_1,\ldots,\eta_4$. 
\begin{equation}
    Q=[\kappa~~\gamma~~\eta_1~~\eta_2~~\eta_3~~\eta_4]
\end{equation}

The mean of a squared defines the objective error fitness function used to quantify the effectiveness of a given controller. It is evaluated online for a step-response of the closed-loop system.

\begin{equation}
    f_{se} = \frac{1}{N} \sum_{k=1}^{N} \Tilde{\theta}^2 +\Dot{\Tilde{\theta}}^2
\end{equation}
with $\Tilde{\theta}=\theta-\theta_d$, $\Dot{\Tilde{\theta}}=\Dot{\theta}-\Dot{\theta}_d$

In order to achieve best-desired performances, nonlinear constraints are considered to overcome the problem of over-taking and static error  \cite{yacef2013pso}.

The constraints are presented as follows

\begin{equation}
\mathcal{C}(\theta) = \left[    
    \begin{array}{c}
         - min_{0\leq t < t_m} \theta(t) - d_0 \\
         max_{0\leq t < t_r} \theta(t) - (O_t\theta_d(t)+\theta_d(t))\\
         - min_{t_m\leq t < t_r} \theta(t) + d_m \theta_d(t) \\ 
         max_{t_r\leq t < t_f} \theta(t) - (\theta_d(t)+e_s\theta_d(t))\\
         - min_{t_r\leq t < t_f} \theta(t) + (\theta_d(t)-e_s\theta_d(t))
    \end{array}
       \right]
\end{equation}
with 
$O_t$ : represent the overtaking;\\
$d_m$ : limited value corresponding increased time $t_m$;\\
$d_0$ : limited value corresponding Started time;\\
$t_r$ : response time;\\
$t_m$ : increased time;\\
$e_s$ : static error.
\begin{figure}[ht]
\centering
\includegraphics[scale= 0.5]{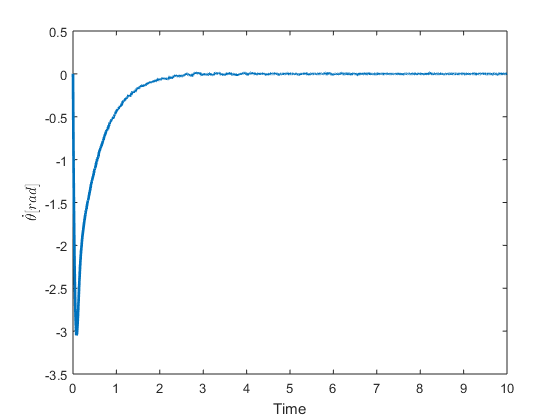}
\caption{Actual and desired  angular velocity of knee joint $\dot{\theta}$}
\end{figure}

\section{SIMULATION RESULTS}
The proposed optimization procedure is used to tune controller parameters (10) such as to minimize the square error sum cost function (11) respecting nonlinear constraints (12) and boundaries limit $ -\frac{\pi}{2}\leq \theta \leq 0 rad$, $-3.1rad/s\leq \dot{\theta} \leq 3.1rad/s$, $-20N\leq \Gamma \leq 20N rad$d for a step response of exoskeleton model. In the simulation, we take the following optimization parameters:

\begin{table}[ht]
\caption{NONLINEAR CONSTRAINTS PARAMETERS}
\label{table_1}
\begin{center}
\begin{tabular}{|c||c|}
\hline
Parameter & Value\\
\hline
$O_t$ & 0.02rad\\
$d_m$ & 0.01rad\\
$d_0$ & 0.01rad\\
$e_s$ & 0.01rad\\
$t_r$ & 1sec\\
$t_m$ & 0.8sec\\
\hline
\end{tabular}
\end{center}
\end{table}

Figure 4 and figure 5 represent respectively step responses of the angular position of knee joint $\theta$ and angular velocity of knee joint $\dot{\theta}$ after many executions using PSO algorithm. Figure 5 represent the delivered torque applied to.\\

Figure 6 represent the delivered torque applied to the system. Figure 6 show the evolution of fitness function. We notice that we have achieved satisfactory results with a static error near to zero, a maximum permissible overtaking, and a response time of about 1 sec. The control laws are bounded. Bounds of optimized parameters are given in table II.

\begin{table}[ht]
\caption{OPTIMISATION PARAMETERS BOUNDS }
\label{table_2}
\begin{center}
\begin{tabular}{|c||c||c|}
\hline
Parameter & Min & Max\\
\hline
$\kappa$ & 1 & 10\\
$\gamma$ & 1 &  5\\
$\eta_1$ & 1 & 15\\
$\eta_2$ & 1 & 15\\
$\eta_3$ & 1 & 15\\
$\eta_4$ & 1 & 15\\
\hline
\end{tabular}
\end{center}
\end{table}

\begin{figure}[ht]
\centering
\includegraphics[scale= 0.5]{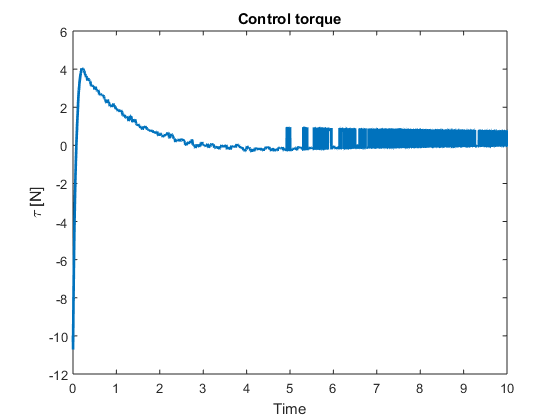}
\caption{Delivered torque $\Gamma$}
\end{figure}

\begin{figure}[ht]
\centering
\includegraphics[scale= 0.5]{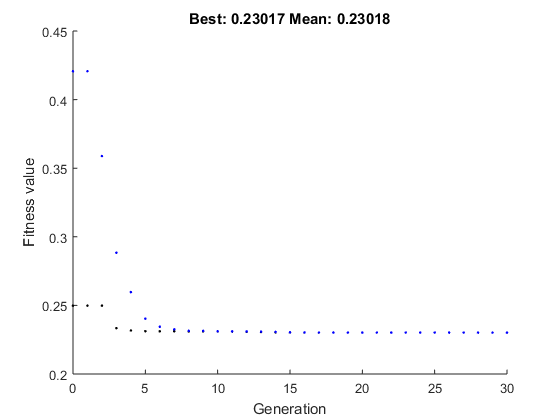}
\caption{Fitness function}
\end{figure}
In figures 7 we can show that the fitness function stabilize in a value equal to 0.23017 in 30 iterations. The tuned controller parameters values are given by table III.

\begin{table}[ht]
\caption{FINAL TUNED PARAMETERS VALUE}
\label{table_3}
\begin{center}
\begin{tabular}{|c||c|}
\hline
Parameter & Final tuned value \\ 
\hline
$\kappa$ & 9.9987 \\
$\gamma$ & 1.0001 \\
$\eta_1$ & 5.3202 \\
$\eta_2$ & 9.9887 \\
$\eta_3$ & 9.6555 \\
$\eta_4$ & 8.0300 \\
\hline
\end{tabular}
\end{center}
\end{table}

\section{CONCLUSIONS}
In this paper, we proposed an optimization procedure based on PSO algorithm with nonlinear constraints to tune an adaptive controller parameters. The proposed method allows to generate optimal torque necessary for rehabilitation of the lower limb. The optimization procedure is based on a squared error fitness function. Some nonlinear constraints are added in the optimization procedure to improve system’s performances.

In future works, we plan to extend the proposed approach to others complicated assistance modes. Also, we expect to validate our approach on real exoskeleton platform.   

\addtolength{\textheight}{-21cm}   


\bibliographystyle{IEEEtran}
\bibliography{referencebiblio}

\end{document}